\documentclass[aps,showpacs,twocolumn]{revtex4}

\usepackage{graphicx}
\usepackage{dcolumn}
\usepackage{bm}
\usepackage{color}

\begin{document}

\title{$p$-wave phase shift and scattering length of $^6$Li}

\author{S. Gautam }
\email{sandeep@prl.res.in}
\author{D. Angom}
\email{angom@prl.res.in}

\affiliation{
   Physical Research Laboratory \\
   Navarangpura, Ahmedabad - 380 009
}

\begin{abstract}
   We have calculated the $p$-wave phase shifts and scattering length of 
   $^6$Li. For this we solve the $p$ partial wave Schr\"odinger equation 
   and analyze the validity of adopting the semiclassical solution to 
   evaluate the constant factors in the solution. Unlike in the $s$ wave 
   case, the semiclassical solution does not provide unique value of the 
   constants. We suggest an approximate analytic solution, which provides  
   reliable results in special cases. Further more, we also use the variable 
   phase method to evaluate the phase shifts. The $p$-wave scattering lengths 
   of $^{132}$Cs and $^{134}$Cs are calculated to validate the schemes 
   followed. Based on our calculations, the value of the $p$ wave scattering 
   length of $^6$Li is $-45a_o$.  
 
\end{abstract}



\maketitle


\section{Introduction}

   For bosonic isotopes of atoms, one parameter which describes the 
low energy scattering properties is $a_0$, the $s$-wave scattering length. 
It arises from the $s$-wave phase shift $\eta_o$. The corresponding 
interaction potential is a crucial parameter in Bose Einstein condensates of 
dilute ultracold atomic gases. The fermionic counterpart is the interaction 
potential arising from the $p$-wave scattering. Calculations of which is 
important as experiments on fermionic isotopes have made impressive strides 
since the first experimental observation of degenerate fermions 
\cite{Demarco}. Superfluidity in two species fermionic mixture of $^6$Li, 
which was first predicted theoretically \cite{Houbiers}, has been observed 
without any ambiguity \cite{Kinast,Bartenstein,Zwierlein}. Following which, 
intense experimental and theoretical investigations continues on the phase 
diagram of the spin polarized two component $^6$Li mixture. Among the most 
recent developments are the theoretical investigation of the phase diagram 
at finite temperature \cite{Parish1} and the experimental investigation 
of the same at unitarity \cite{Yong}. Further, the recent 
achievement of cooling the $^6$Li-$^{40}$K fermionic mixture to degeneracy 
\cite{Taglieber} takes us closer to observing exotic phases predicted for 
spin polarized heteronuclear fermionic mixtures \cite{Parish2}.  Some of the
predicted phases are fragile and crucially dependent on the difference 
of the chemical potentials. For such cases, the change in chemical 
potential induced by the $p$-wave scattering is likely to be an important 
parameter. The pseudopotentials arising from the higher partial waves,
discussed in recent works \cite{Idziaszek,Kanjilal}, can be used to 
incorporate the effects of $p$-wave scattering in fermionic isotopes.  

  In this paper we describe the calculation of the $p$-wave scattering 
length of $^6$Li. As suggested in an earlier work on $s$-wave scattering 
\cite{Gribakin}, the WKB  method is used to determine the constant 
parameters of the partial wave solutions of the Schr\"odinger equation. 
However, unlike in the $s$-wave scattering, the parameters in the $p$-wave 
calculations has radial dependence. This is an outcome of including the 
centrifugal potential in the effective interatomic potential. To circumvent 
this, an analytic expression, which provide an estimate of the $p$-wave 
phase shift is suggested. This method is valid when the dispersion constant 
$C_6$ is large. We also calculate the phase shift using the variable phase 
method. To test and validate the numerical schemes adopted, the $s$-wave 
phase shift of $^{133}$Cs and $p$-wave scattering length of $^{132}$Cs and 
$^{134}$Cs are calculated.

  For completeness in Section.I of the paper, we provide an outline of
solving the $p$ partial wave Schr\"odinger equation in different radial
regions. The Section.II discusses the calculation of the scattering length
using the WKB method to determine the constants in the partial wave 
solutions.  Then a brief description of the variable phase method is 
provided in Section.III, this is followed with results and conclusions.  All
the equations and results in this paper are in atomic units, in which
$\hbar=m_e=e=1$.


\section{$p$-wave phase shift}
The radial part of the Schrodinger wave equation for collisions
between two atoms with $V(R)$ as the interatomic potential is
\begin{equation}
  \label{eq.a.a}
  \frac{d^2\chi}{dR^2} + \left[k^2 -U(R) -\frac{l(l+1)}{R^2} \right]\chi
   =0,
\end{equation}
Here $U(R)=2mV(R)$, $k$ is the relative momentum of the two atoms,
$l$ is the angular momentum quantum number and $\chi(R)$ is the radial
wave-function. This equation can be used to calculate scattering
phase shifts ($\eta_l$) which each component angular momentum suffers
due to the interaction with the scattering center.
At low energies we can calculate approximate solution of Eq.~(\ref{eq.a.a})
in three difference radial ranges. These are the $ kR\ll1$, $kR<1$ and
$kR\approx 1$ regions.


\subsection{$kR\ll1$ region}
\label{seciia}

In this region, we can neglect the $k^2$ term
\begin{equation}
   \label{eq.a.b}
   \frac{d^2\chi}{dR^2} + \left[ -U(R) - \frac{l(l+1)}{R^2} \right]\chi
    = 0,
\end{equation}
and hence, the solution is independent of $k$. At large distances from the
scattering center, van der Waal's potential
$U(R)=-2m\alpha/(R^6)=-\gamma^2/R^6$ is the dominant inter-atomic
interaction. Here, $\alpha$ is the van der Waal's coefficient. For low energy
collisions $k\rightarrow 0$, the interatomic potential approaches the
asymptotic form within the $kR\ll 1$ region. Then,
\begin{equation}
  \label{eq.a.h}
  \frac{d^2\chi}{dR^2} +\left[ \frac{\gamma^2}{R^6}-
  \frac{l(l+1)}{R^2}\right]\chi =0.
\end{equation}
Substituting $\chi(R)=\xi(R)\sqrt{R}$ and $R=\sqrt{\gamma/2x}$, we  obtain
the Bessel differential equation of order $(2l+1)/4$
$$
  \frac{d^2\xi}{dx^2} + \left[ 1- \frac{(2l+1)^2}{16x^2}\right]
  \frac{d\xi}{xdx}= 0.
$$
The general solution of Eq.(\ref{eq.a.h}) is
$$
  \chi(R) =\sqrt{R}\left[ AJ_{(2l+1)/4}\left(\frac{\gamma}{2R^2}\right)+
        B J_{-(2l+1)/4}\left(\frac{\gamma}{2R^2}\right)\right].
$$
Considering the $p (l=1)$ partial wave
\begin{equation}
  \label{eq.a.l}
  \chi(R) = \sqrt{R} \left[A J_{3/4}\left(\frac{\gamma}{2R^2}\right) +
         B J_{-3/4}\left(\frac{\gamma}{2R^2}\right)\right].
\end{equation}
In the limit $R\rightarrow\infty$ or $x\rightarrow0$, where $kR< 1$, the
leading terms in the series expansion of  $J_{3/4}(x)$ and $J_{-3/4}(x)$ are
dominant and the remaining terms are negligible
\begin{equation}
  \label{eq.a.m}
  \chi(R) = B \frac{(\gamma/2)^{-3/4}R^2}{2^{-3/4}\Gamma(1/4)}
             + A \frac{(\gamma/2)^{3/4}}{2^{3/4}\Gamma(7/4)R} ,
\end{equation}
where $\Gamma (\ldots)$ are the gamma functions.


\subsection{$kR < 1$ region}
 In this region, we can neglect the interatomic interaction potential $U(R)$,
if it approaches zero faster than $1/R^2$. In addition, as in the $kR \ll 1$
region, the $k^2$ term  can be neglected
\begin{equation}
  \label{eq.a.c}
  \frac{d^2\chi}{dR^2} -\frac{l(l+1)}{R^2}\chi =0.
\end{equation}
The most general solution of this equation is
\begin{equation}
  \label{eq.a.d}
  \chi(R) = aR^{l+1} + bR^{-l}.
\end{equation}
For $l=1$, comparing the solutions in Eq.~(\ref{eq.a.m}) and 
Eq.~(\ref{eq.a.d})
\begin{equation}
  \label{eq.a.n}
  a = B \frac{(\gamma/2)^{-3/4}}{2^{-3/4}\Gamma(1/4)}
  \;\;\;\; \mbox{and}\;\;\;\;
  b = A \frac{(\gamma/2)^{3/4}}{2^{3/4}\Gamma(7/4)}.
\end{equation}
The comparison is possible as $\chi$ in the $R\rightarrow \infty$ of 
Eq.(\ref{eq.a.l}) has the same form as the solution in the $kR < 1$ region.


\subsection{$kR \approx 1$ region}
Increasing $R$ further, we enter the $kR\approx 1$ region. In this region,
we can neglect $U(R)$ in comparison to the other terms in Eq.~(\ref{eq.a.a}),
then
\begin{equation}
  \label{eq.a.e}
  \frac{d^2\chi}{dR^2} + \left[ k^2 -\frac{l(l+1)}{R^2}\right] \chi =0.
\end{equation}
This is the Schr\"odinger equation in the asymptotic region and the solution
is a plane wave of wave number $k$. However, the interatomic potential
introduces a phase shift ($\eta _l$)\cite{Mott} to the plane wave
$$
  \label{eq.a.f}
  \chi(R) = CR^{1/2}\left[J_{l+1/2}(kR) + (-1)^l \tan(\eta_l)J_{-l-1/2}(kR)
         \right].
$$
Expanding the Bessel's functions $J_{l+1/2}(kR)$ and $J_{-l-1/2}(kR)$,
the solution when $kR < 1$  is
\begin{eqnarray}
  \label{eq.a.g}
  \chi(R)& = &\left(\frac{2k}{\pi}\right)^{1/2}C\left[R^{l+1}\frac{(2k)^l
             \Gamma(l+1)}{\Gamma(2l+2)} +  \right . \nonumber \\
         && \left . R^{-l}\tan{\eta_l}\frac{(2k)^{-l-1}2\Gamma(2l+1)}
            {\Gamma(l+1)}\right].
\end{eqnarray}
Equating the solutions in Eq.~(\ref{eq.a.d}) and ~(\ref{eq.a.g})
\begin{equation}
  \label{eq.b.a}
  \tan~\eta_l=\frac{b 2^{2l} (\Gamma(l+1))^2}{a
              \Gamma(2l+1) \Gamma(2l+2)}k^{2l+1} .
\end{equation}
Since $b/a$ is independent of $k$, $\tan\eta_l$ varies as $k^{2l+1}$. It
should also be mentioned that, though the entire radial range is divided
into three domains, the solutions are in the asymptotic region. This defines
the phase shift $\eta_l$ in terms of the $b/a$ and it is strictly
applicable for $k\rightarrow 0$. For the $p$-partial wave
\begin{equation}
  \label{eq.c.a}
  \tan~\eta_1=\frac{b}{a}\frac{4}{\Gamma(3) \Gamma(4)}k^3.
\end{equation}
In the above expression, to calculate $\eta_1$, the ratio $b/a$ should be
determined. From the Eq.(\ref{eq.a.n}), the phase shift can also be defined
in terms of $A/B$ as
\begin{equation}
  \label{eq.c.b}
  \tan~\eta_1=\frac{1}{24}\frac{A}{B}
              \frac{\gamma ^{3/2}\Gamma(1/4)}{\Gamma(7/4)} k^3 .
\end{equation}
Here the actual values of $\Gamma(3)$ and $\Gamma(4)$ are used. Now onwards
this definition of $\eta_1$ is used.  As the phase shift $\eta_1$ has
functional dependence on $\tan ^{-1}$, which has a domain of 
$[-\infty, \infty]$, its accuracy is sensitive to the value of $A/B$.


\section{Scattering length calculation }
As mentioned earlier, the expression of $\eta_1$ in Eq.(\ref{eq.c.b}) is in
the asymptotic domain of the interatomic potential and require evaluation of
$A/B$. The solution of the Schr\"odinger equation Eq.(\ref{eq.a.a}) in the
asymptotic region, given by Eq.(\ref{eq.a.l}) and parametrized in terms of $A$
and $B$, is the outer solution. To determine $A/B$, the Eq.(\ref{eq.a.a})
should be solved within the inner part of the interatomic potential: inner
wall and well. This can be evaluated using the WKB method \cite{Gribakin}.
Matching the solutions, outer and inner, at a point we can evaluate $A/B$.
The matching point should be in the region where the asymptotic form of the
interatomic potential begins to dominate but the WKB solution is still valid.

\subsection{WKB solution}
\begin{figure}
   \includegraphics[width=8cm]{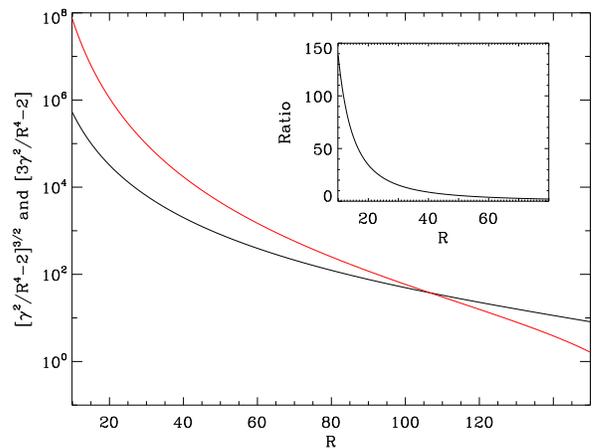}
   \caption{Terms in the relation $[3\gamma^2/R^4-2]\ll[
            \gamma^2/R^4-2]^{3/2}$ which defines the validity
            of the WKB solution are plotted for Cesium. The red and black 
            curves correspond to $[\gamma^2/R^4-2]^{3/2}$  and 
            $[ 3\gamma^2/R^4-2]$ respectively. Inset shows the ratio of 
            the two terms. }
  \label{fig.1}
\end{figure}

For partial waves other than $s$, the centrifugal potential is
nonzero. Combining the  the interatomic potential and centrifugal term, the
effective interatomic potential
\begin{equation}
   U_{\rm eff}(R) = U(R)+l(l+1)/R^2 .
\end{equation}
Then the local momentum of the scattered atom $p(R)= \sqrt{-U_{\rm eff}(R)}$,
and the de Broglie wavelength $\lambda (R) = 2\pi/\sqrt{-U_{\rm eff}(R)}$.
The  WKB approximation is applicable when
\begin{equation}
  \label{eq.a.p}
  \left |\frac{d\lambda}{dR}\right|\ll2\pi\;\;\mbox{or} \;\;
   \frac{m\vert F \vert}{p^3}\ll1 .
\end{equation}
In above relation $F=-dU_{\rm eff}/(2mdR)$ is the net force acting on the atom.
At large distances, when the interatomic potential approaches the asymptotic
form, the above inequality for $l=1$ is
\begin{equation}
  \label{eq.a.r}
  \left[ \frac{3\gamma^2}{R^4}-2\right]\ll\left[
       \frac{\gamma^2}{R^4}-2\right]^{3/2} .
\end{equation}
This inequality is in general valid up to large radial distances. As an
example Fig.\ref{fig.1} shows that for Cs the inequality is satisfied
up to $\approx 30$. Then the WKB solution at $R$ which satisfies the 
inequality and larger then the classical turning point $R_o$ is
\begin{equation}
  \label{eq.a.q}
  \chi(R) =\frac{c}{\sqrt{p}}\cos\left(\int_{R_o}^R p dR -
           \frac{\pi}{4}\right)
\end{equation}
and the logarithmic derivative is
\begin{eqnarray}
 \!\!\!\!\!\!\! \zeta(R) &= &\frac{\chi'(R)}{\chi(R)} \nonumber \\
            &= &-p(R)\tan \left (\int_{R_o}^RpdR
              -\frac{\pi}{4} \right ) -\frac{1}{2p}\frac{dp}{dR}.
   \label{zeta}
\end{eqnarray}
An analytic expression of $\zeta(R)$ for $l=1$ is obtained if only the van 
der Waal's potential is considered
\begin{eqnarray}
 \label{eq.d.b}
 \zeta(R)&=& -\frac{\sqrt{\gamma^2-2 R^4}}{R^3}
        \tan \left[ \frac{\sqrt{\gamma^2- 2 R_o^4}}{2R_o^2}-
        \frac{\sqrt{\gamma^2-2 R^4}}{2R^2} \right. \nonumber \\
     &&-\left. \frac{1}{\sqrt{2}}\left( \sin^{-1}{
        \frac{\sqrt2 R^2}{\gamma}}- \sin^{-1}\frac{\sqrt 2 R_o^2}{\gamma}
        \right) - \frac{\pi}{4} \right ]\nonumber \\
     &&- \left[ \frac{2R^4-3\gamma^2}{2R(\gamma^2-2R^4 ) }\right].
\end{eqnarray}
This approximation neglects the well and inner wall part of the
interatomic potential. It is appropriate for model potentials comprising
of the van der Waal's and hard core potential. The above
expression is reduced to that of $s$ case when $l=0$, which is used
implicitly in ref. \cite{Gribakin}.

  From the inequality (\ref{eq.a.r}), it is evident that when $\gamma$ is
sufficiently large, WKB solution is valid up to radial distances where the
asymptotic form of the potential begins to dominate. Choose a point $R^*$ 
in this region, such that the $2$ in the inequality can be neglected. Then 
the inequality is modified to
\begin{equation}
  \label{eq.d.a}
  R^*\ll(\gamma /3)^{1/2} .
\end{equation}
At this point, as the interatomic potential has the asymptotic form,
Eq.~(\ref{eq.a.l}) is a solution of the Schr\"odinger equation
Eq.(\ref{eq.a.a}). This is discussed in Section.\ref{seciia}.


\subsection{Matching $\chi'(R)/\chi(R)$ }
 From the definition,  at $R^*$ and around it $x = \gamma /(2R^{*2})\gg 1$.
Where $x$ is the variable first used in Section \ref{seciia}. We can then
use the asymptotic expression expression of the Bessel's function
$$
  \label{eq.a.t}
  \lim_{x\rightarrow \infty} J_n(x) \simeq \sqrt{\frac{2}{\pi x}} \cos
      \left[ x-\left( n + \frac{1}{2}\right) \right],
$$
and the solution in Eq.~(\ref{eq.a.l}) is
\begin{eqnarray}
  \label{eq.a.u}
 \chi(R) &= &\frac{2R^{3/2}}{\sqrt{\pi\gamma}}\left[ A~\cos\left(
           \frac{\gamma}{2R^2}- \frac{5\pi}{8}\right)\right . + \nonumber \\
         && \left .B~\cos \left( \frac{\gamma}{2R^2}+\frac{\pi}{8}\right)
            \right].
\end{eqnarray}
This is the inner limit of the asymptotic solution. Whereas the
solution in Eq.(\ref{eq.a.m}), in the $x\rightarrow 0$, is the outer limit of
the asymptotic solution. To determine $A/B$ we match the logarithmic 
derivatives of the analytical solution in Eq.~(\ref{eq.a.u}) and WKB solution
in Eq.~(\ref{eq.a.q}). The ratio is
\begin{equation}
  \label{eq.a.v}
  \frac{A}{B} = \frac{\sqrt{2m\alpha}\sin{\phi_2}- R^{*2}[R^*\zeta(R^*)-
                3/2]\cos{\phi_2}}{R^{*2}[R^*\zeta(R^*)-3/2]\cos{\phi_1}-
                \sqrt{2m\alpha}\sin{\phi_1} },
\end{equation}
here $\phi_1=\gamma/(2R^{*2}-5\pi/8)$, $\phi_2=\gamma/2R^{*2}+\pi/8$. It is 
evident from the above expression that $A/B$ depends on the choice of the
matching point $R^*$. This implies that $\eta_1$ and hence the scattering
length, to be defined later, depend on $R^*$. This radial dependence arises
from the centrifugal part in the expression of the effective potential
$U_{\rm eff}(R)$. Neglecting the terms arising from the centrifugal
potential in the expression of $\zeta(R)$, logarithmic derivative 
of the WKB solution is
\begin{equation}
\label{eq.Z.a}
  \zeta(R)=-\frac{\gamma}{R^3}\tan\left( \frac{\gamma}{2}
            \left(\frac{R^2-R_o^2}{R^2R_o^2} \right)-\frac{\pi}{4} \right)+
            \frac{3}{2R}.
\end{equation}
For the obvious reason, large value of $R/\gamma$, this expression is not
valid at very large values of $R$. Assuming that the maximum radial distance 
at which the above expression is valid, the interatomic potential has the
asymptotic form. If  $A/B$ is calculated using Eq.~(\ref{eq.Z.a}) as the WKB 
solution, it is then independent of the choice of matching point
\begin{equation}
\label{eq.Z.b}
\frac{A}{B}= -\frac{\sin\left( \phi-\frac{\pi}{8}\right) }
              {\sin\left(\phi-\frac{7\pi}{8} \right) },
\end{equation}
where $\phi=\gamma/(2R_o^2)$. This is however a crude estimate as the
outer solution includes the effect of centrifugal term but the inner solution,
the WKB solution, does not. Still, it is a useful estimate, as it is
an analytic expression and not difficult to evaluate.


\subsection{$p$-wave scattering length}

For any arbitrary potential which vary as $R^{-n}$, the scattering length
($a_l$) and phase shift ($\eta_l$) are related as \cite{Mott}
\begin{equation}
  \label{eq.a.z}
  \lim_{k \to 0}~k^{2l+1}\cot~\eta_l = \frac{-1}{a_l^{2l+1}}.
\end{equation}
This relation is valid provided $l<(n-3)/2$ and is applicable for $p$
partial wave when $n=6$ or higher. From Eq.~(\ref{eq.c.b}) and (\ref{eq.a.z})
scattering length of the $p$ partial wave is
\begin{equation}
  \label{eq.b.b}
  a_1^3=-\frac{1}{24}\frac{A}{B}\frac{\gamma^{3/2}\Gamma(1/4)}{\Gamma(7/4)}.
\end{equation}
Using the value of $A/B$ in Eq.(\ref{eq.a.v}), the $p$-wave scattering length
$a_1$  can be evaluated. A rough estimate of $a_1$ is to use the analytical
expression in Eq.(\ref{eq.Z.b}). The quantity $a_1^3$ is referred to as 
the $p$wave scattering volume. It is an important parameter in 
the three-body recombination of identical, spin polarized fermions 
in low tempertures \cite{Suno}.


\section{Variable phase method}
 Another approach to calculate $a_1$ is to evaluate $\eta_1$ from the phase
function equation, which is referred to as the variable phase method
\cite{Calogero}. The method is applicable when the interatomic potential
satisfies two conditions. First condition is, the interatomic potential
should be less singular at origin than the centrifugal part, that is
\begin{equation}
 \lim_{R\rightarrow0}[R^2U(R)]=0 .
\end{equation}
This implies that, near the origin $U(R)\rightarrow U_0R^m$  as
$R\rightarrow 0$, with $m>-2$, where $U_0$ is a constant. The interatomic
potential satisfies this condition since at smaller radial distances, it
approaches the inner wall and is repulsive.

\begin{figure} [h]
  \includegraphics[width=8cm]{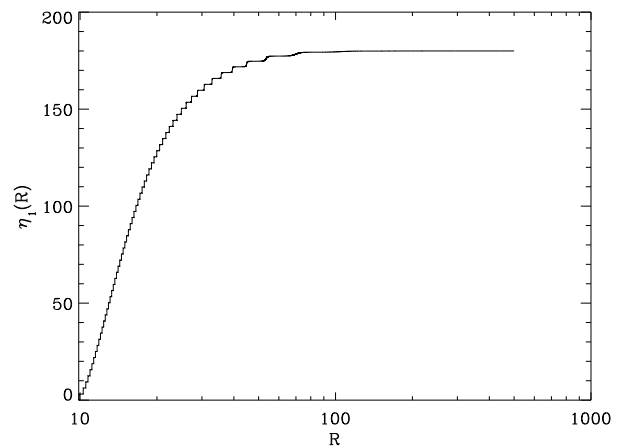}
  \caption{The phase function $\eta_1(R)$ for $^{134}$Cs at $k=0.05$ in atomic
           units. The radial distance is in $\log$ scale and the plot shows
           $\eta_1(R)$ up to $R=500 a_0$, where the the phase function
           converges to an accuracy of $10^{-6}$. }
  \label{fig.ps}
\end{figure}
The second condition is, the interatomic potential should decrease faster
than the Coulomb potential or inverse of $R$. In the variable phase method,
the phase shift $\eta_l$ is a solution of the nonlinear differential
equation  \cite{Calogero}
\begin{equation}
\label{eq.ps.a}
 \eta_l'(R)=-k^{-1}U(R)[\cos\eta_l(R)j_l(kR)-\sin\eta_l(R)n_l(kR)]^2,
\end{equation}
where $j_l(kR)$ and $n_l(kR)$ are the spherical Bessel and Neumann
functions respectively. The above equation can be used to evaluate the
phase function $\eta_l(R)$, which defines the total phase shift upto $R$. 
Then the phase shift is the limiting value
\begin{equation}
 \label{eq.ps.b}
 \eta_l= \lim_{R\rightarrow \infty}\eta_l(R).
\end{equation}
To calculate $\eta_l$, the phase function $\eta_l(R)$ is evaluated upto a 
cut off point at which $\eta_l(R)$ saturates. The physical implication of
the cut off point is, it is the radial distance beyond which  the inter
atomic potential can be considered zero. To illustrate the radial dependence
and saturation Fig.\ref{fig.ps} shows  $\eta_1(R)$ for $k=0.05$ of 
$^{134}$Cs. The figure shows that, for the chosen parameters, saturation 
occurs at $\approx 500$. The phase shift $\eta _l$, as it is evident from 
Eq.~(\ref{eq.ps.a}), depends on the relative momentum of the 
colliding atoms $k$. To remove the mod($\pi$) ambiguity, the phase shift 
is normalized such that it approaches zero as $k\rightarrow \infty$, that is
\begin{equation}
 \lim_{k\rightarrow \infty}\eta_l=0.
\end{equation}
In addition to above condition $\eta_l$ is considered to be regular function
of $k$. The Eq.(\ref{eq.ps.a}) is numerically integrated from the classical
turning point, where the inner wall starts, to the cut off point.
The solution is the phase shift as defined in Eq.(\ref{eq.ps.b}).


\section{Results}
To evaluate the phase shift and the scattering length, the interatomic
potential should be known accurately. In this paper, we present the
results of our calculations for cesium and lithium atoms. For which the
interatomic potentials are known accurately and hyperfine interactions are 
less significant in the long range part of the potentials. The cesium 
interatomic potential is the one used in the work of Gribakin and Flambaum 
\cite{Gribakin}, which is based on ref \cite{Radstig}. The lithium 
interatomic potential is based on the work of Zemke and Stwalley 
\cite{Zemke93}. The same authors and their collaborators have also 
calculated the interatomic potential of  other alkali metals sodium 
\cite{Zemke94a} and potassium \cite{Zemke94b}. However, these are
limited to radial distances where hyperfine interactions are not important.

 Once the interatomic potential is known, the logarithmic derivative of the
WKB solution $\zeta(R)$, given in Eq.(\ref{zeta}) is evaluated
numerically. To check the accuracy and validate the numerical schemes 
adopted, the numerically calculated $\zeta(R)$ with only the van der Waal's 
potential is compared with the analytic expression in Eq.(\ref{eq.d.b}). 
The calculated $\zeta(R)$ is then used in Eq.(\ref{eq.a.v}) 
to evaluate $A/B$. As discussed earlier, the calculated $A/B$  has radial
dependence and it is difficult to choose an appropriate matching point $R^*$.
An  estimate of $R^*$ can however be obtained from the 
results of the variable phase calculations. Such a calculation provides a 
consistency check on the use of the WKB method. Despite the radial 
dependence, for the $p$-wave calculations the importance of using 
the WKB solution lies in Eq.(\ref{eq.Z.b}), an analytic expression of $A/B$. 
From which, it is possible to calculate a rough estimate of the scattering 
length, which is in reasonable agreement with the numerical
result for $^{134}$Cs. For $^{132}$Cs and $^6$Li, the two results are very 
different. 

 The phase shift $\eta_1$ is also calculated from the phase function
equation, for which the nonlinear differential equation Eq.~(\ref{eq.ps.a}) 
is numerically solved . In the present work,
the Runge-Kutta-Fehlberg (RKF) method is used. With this method, it is
possible to calculate phase shifts for small values of $k$, which is
otherwise not possible with methods  like fourth order Runge-Kutta.
Calculations with the later method has large errors for small values of $k$.
For example, in  the Cesium calculations with RK4 method, the calculated
phase shifts  are reliable up to $k=6.0\times 10^{-3}$  and
$k=1.1\times 10^{-3}$  for $p$ and $s$ partial waves respectively. In
comparison, with RKF method one can calculate phase shifts for still lower
values of $k$. In the numerical calculations, to integrate the phase 
function differential equation, the classical turning point
is chosen as the starting point. Then, zero is an appropriate initial value
of  $\eta_1(R)$. Using RKF we can calculate $\eta_1(R)$ for different values
of $k$. The $\eta_1(R)$ is calculated till a point where it does not increase
with further increase in $R$. This point is chosen as the cut off point where
potential can be considered as zero and the corresponding value of the
phase function is the required phase shift. To calculate the scattering 
length $a_1$, the phase shift $\eta_1$ is evaluated for a range of $k$ close
to zero. This is essential as $a_1$ depends on the nature of the $\tan\eta_1$
close to $k=0$. From the definition in Eq.~(\ref{eq.a.z}),
$-\tan\eta_1/k^3$  converges to the scattering length for $p$ partial
wave as $k\rightarrow 0$. Alternatively, $-\tan\eta_1/k^2$ is linear
as $k\rightarrow 0$ and the slope is the scattering length.


\subsection{Cesium}
Consider the scattering of two Cesium atoms in the $^3\Sigma_u$ state . The
form of interaction potential is \cite{Radstig}
\begin{equation}
  \label{eq.pot}
  V(R)=\frac{1}{2}BR^\alpha\exp{\beta R}-\left(\frac{C_6}{R^6}+
              \frac{C_8}{R^8}+\frac{C_{10}}{R^{10}} \right)f_c(R) .
\end{equation}
where $B=0.0016$, $\alpha = 5.53$ and $\beta=1.072$. These are given in
ref \cite{Gribakin} and $f_c(R)$ is the cut-off function and has the
expression
\begin{equation}
   f_c(R)=\Theta(R-R_c) + \Theta(R_c-R)\exp^{-(R_c/R-1)^2},
\end{equation}
where $\Theta(\ldots)$  is the step function which is equal to $1$ or $0$,
depending on whether its argument is greater or less than zero. Here $R_c$ is
the cut off parameter.
\begin{figure}
   \includegraphics[width=8cm]{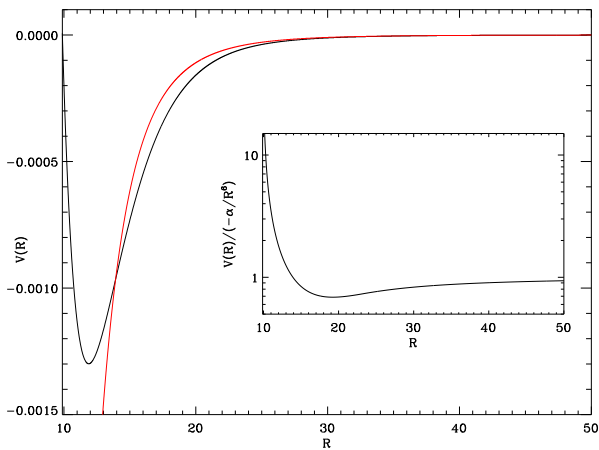}
   \caption{The black and red curves are the plots of interatomic and the van
            der Waal's potential respectively for Cesium. In the inset ratio
            of van der Waals term to the actual potential is plotted. }
   \label{ces_pot}
\end{figure}
For comparison, the total interatomic potential $V(R)$ and van der Waal's
potential are shown in Fig.\ref{ces_pot}. The ratio of the two are also
shown as an inset plot. The figure shows that $V(R)$ approaches the 
asymptotic form at around $R=24$, where the ratio of the two is $0.8$. 
Beyond $R=30$, the two are indistinguishable. Using this potential,
the phase shift and scattering lengths are calculated for the $^{132}$Cs 
and $^{134}$Cs fermionic isotopes. As test calculations, we evaluated the 
$s$ wave phase shifts for the $^{133}$Cs isotope. It is to be mentioned 
that, the value of $\gamma$ is 41240.1. However, in ref. \cite{Gribakin} 
the value of $\gamma$ is defined as 41200. Using the later, the results 
of the test calculations are in good agreement with that of ref. 
\cite{Gribakin}.

 From Eq~.(\ref{eq.Z.b}), which provides a rough estimate, the value of 
$a_1$ is $137$a.u and $-108$ a.u.  for $^{132}$Cs and $^{132}$Cs 
respectively.
\begin{figure}[h]
   \includegraphics[width=8cm]{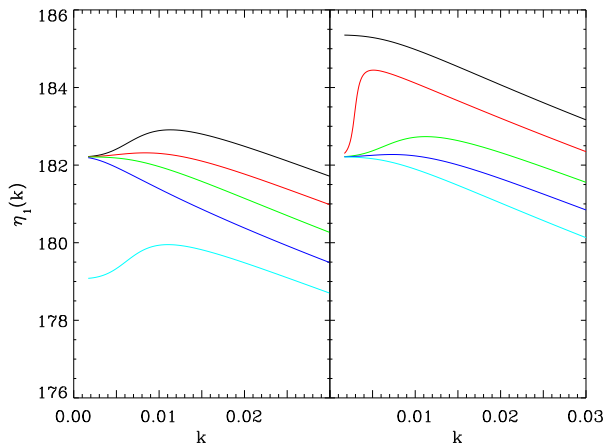}
   \caption{The left and right panel of the figure are the phase shift 
            $\eta_1(k)$ of $^{132}$Cs and $^{134}$Cs isotopes respectively.
            Each panel show the plots of $\eta_1(k)$ for different cut off
            radius $R_c$ of the potential. The, black, red, green, blue and
            cyan correspond to $R_c=23.115$, $23.140$, $23.165$, $23.190$
            and $23.215$ respectively. 
            }
   \label{cseta}
\end{figure}
The $\eta_1$ calculated from the variable phase method, for different
$R_c$ for a range of $k$ close to zero are shown in Fig.\ref{cseta}. 
Our calculations show that, the value of $\eta_1$ at higher values
of $k$ are sensitive to $\gamma$, the dependence is nonlinear as the
equation of phase function is a nonlinear differential equation.
Consequently, small change in $\gamma$ could result in significant change 
of $\eta_1$. The dependence $\eta_1$ on $\gamma$ for larger difference
is evident from Fig.\ref{cseta}, which shows $\eta_1$ for $^{132}$Cs and 
$^{134}$Cs.  The difference in the interatomic potential of the two isotopes 
is the mass, which manifests as unequal values of $\gamma$. There are two
distinct differences in the results for the two isotopes: for the 
$R_c=23.215$ case $\eta_1$ converges to $179.08$ and $182.21$ for
$^{132}$Cs and $^{134}$Cs respectively; and for the $R_c=23.115$ case
$\eta_1$ converges to $182.22$ and $185.35$ respectively. The phase 
function in the neighborhood of $k=0$ is also sensitive to the 
accuracy of the integration. 

In the present calculations, the phase function equation is integrated till
$\eta_1$ converges to the order of $10^{-6}$, which is consistent with the
choice of tolerance. For Cs, this requires integration up to radial distances
of $\sim 500$. The large radial distance is necessary as the phase shift
is an asymptotic property. In addition, for the $p$ wave scattering the
centrifugal potential, which has $R^{-2}$ dependence, vanishes at a slower
rate compared to the van der Waal's potential at large radial distances. This
slows the convergences of the phase shift. A calculation of the $s$ wave
phase shift confirms this. The $s$ wave phase shift $\eta_0$ converges to
the order of $10^{-6}$ when the phase equation is integrated to a radial
distance of $\approx 100$.

\begin{figure} [h]
   \includegraphics[width=8cm]{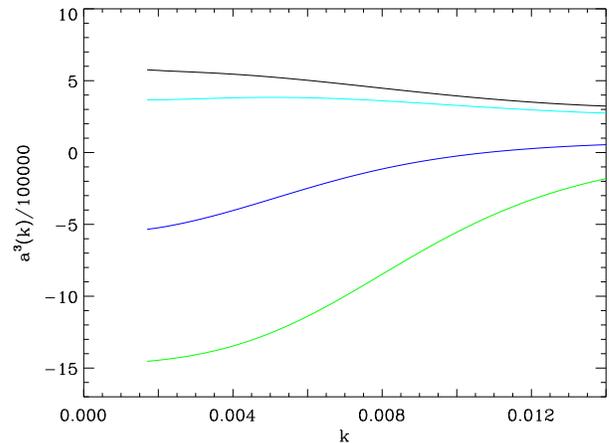}
   \caption{Cube of scattering length $a_1^3(k)$ or -$\tan\eta_1/k^3$ for
            $^{134}$Cs is shown as a function of $k$. The black, green, blue 
            and cyan correspond to  $R_c=23.115$, $23.165$, $23.190$
            and $23.215$ respectively.}
   \label{csa}
\end{figure}
The values of  $-\tan(\eta_1(k))/k^3$ for $^{134}$Cs within a range of $k$ 
for different $R_c$ are shown in Fig.\ref{csa}. As discussed in ref. 
\cite{Gribakin}, the $R_c=23.165$ is a physically reasonable choice. For this 
the scattering length calculated as the slope of $-\tan(\eta_1)/k^2$ in the 
$k\rightarrow 0$ are $53$ and $-113$ in atomic units for $^{132}$Cs and
$^{134}$Cs respectively. This differs from the analytic values mentioned 
earlier by 61\% and 4.4\% for $^{132}$Cs and $^{134}$Cs respectively. 
It shows that the analytic expression though crude provides a
good estimate for $^{134}$Cs, which has higher $\gamma$ .

\begin{figure}[h]
  \includegraphics[width=8cm]{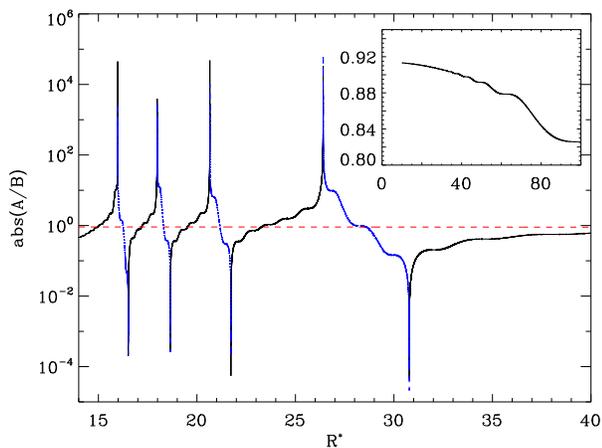}
  \caption{$|A/B|$ for $^{134}$Cs evaluated from the WKB solution is shown 
           as a function of the matching point $R^*$. The black and blue 
           portions of the curve indicate the positive and negative values 
           of $A/B$. The dashed red curve shows the $A/B$ when only the 
           van der Waal's penitential is considered and the inset plot 
           shows the overall nature of the same.
           }
  \label{fig.abyb}
\end{figure}

The evaluation of $\eta_1$ from the numerically calculated $A/B$ poses
some difficulty. This is to do with the choice of appropriate $R^*$,
the matching point, the dependence is shown in the Fig.\ref{fig.abyb}.
The over all trend of the variation is rather complicated for the actual
interatomic potential but less when only the van der Waal's potential
is considered. However, it is possible to calculate $A/B$ from the
value of $a_1$ calculated earlier, which is obtained from the variable
phase method. The value of $A/B$ obtained from such a calculation are
$\approx -0.109$ and $\approx 1.042$ for $^{132}$Cs and $^{134}$Cs 
respectively.
The value of $R^*$ around which $A/B$ is close to this value occurs around
the radial range of $23-24$ atomic units. This is not surprising, as 
mentioned earlier, this is the radial range where the interatomic potential 
approaches the asymptotic form.


\subsection{Lithium}
\begin{figure}[h]
   \includegraphics[width=8cm]{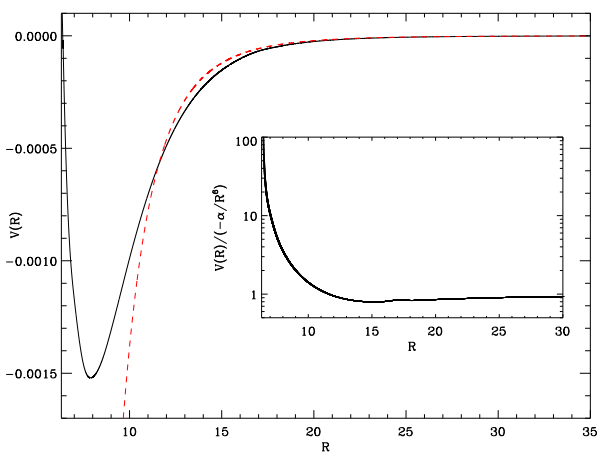}
   \caption{Interatomic potential for $a\Sigma_u^+$ of $^6 \rm{Li}_2$. The
            solid (black) and dash (red) are the interatomic potential and
            the van der Waal's potential respectively. The plot in the inset
            is the ratio of the van der Waal's potential and the interatomic
            potential.}
   \label{lith_pot}
\end{figure}

For the  $a^3\Sigma^+_u$ state of $^6 \rm{Li}_2$, we use the interatomic 
potential suggested by Zemke and Stwalley \cite{Zemke93}. The long range 
part of the  interaction potential, for $^7$Li in particular, are discussed
in ref. \cite{Cote}. For the $R\leq6.388674$ and $R\geq18.0$ regions,
the analytic expressions recommended in ref. \cite{Zemke93} are used.  Then,
for the $6.388674<R<18.0$ region, the interatomic potential is approximated 
as the cubic spline fitted to the potential values given by Zemke and 
Stwalley. A plot of the  interatomic potential obtained is shown in 
Fig.\ref{lith_pot}.
As to be expected, the potential approaches the asymptotic form, van der 
Waal's potential, at smaller radial distance compared to Cs. This is evident
from a visual comparison of the two potentials shown in Fig.\ref{lith_pot}
and Fig.\ref{ces_pot}. This implies that, the phase shift calculation of 
lithium can have better convergence properties. Consequently, the methods 
adopted for the Cs calculations are applicable to Li and provide results with
higher accuracy.

\begin{figure}[h]
   \includegraphics[width=8cm]{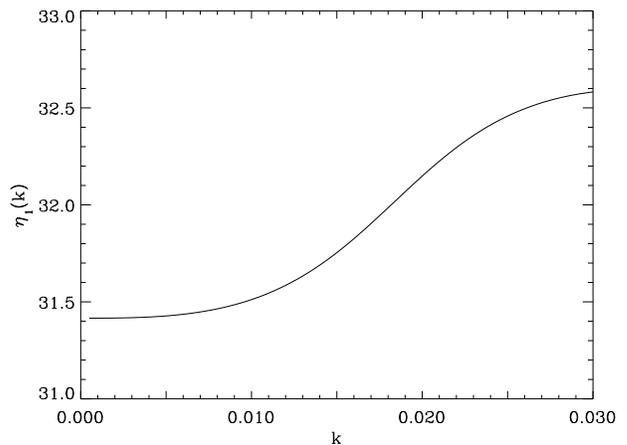}
   \caption{The $p$-partial wave phase shift $\eta_1(k)$ for the 
            $^3\Sigma^+_u$ interaction potential of $^6\rm{Li}_2$. The phase 
            shift converges to 31.46 as $k\rightarrow 0$.}
\label{li_eta}
\end{figure}

 To optimize the calculations, we use the RKF method to solve 
Eq.(\ref{eq.ps.a}) with relative tolerance varied as a function of $k$. 
The relative tolerance is set to $10^{-5}$ for $k>0.01$ and it is
lowered as $k$ is decreased. This is essential as the scattering length 
depends on the nature of the phase shift in the neighborhood of $k=0$. 
Based on extensive test calculations, the optimal choice of the relative
tolerance is $10^{-6}$ for $0.01 \leq k<0.005$ and $10^{-7}$ for 
$k\leq0.005$. With this choice it is possible to get reliable results for 
phase shift up to $k=0.0005$. We find that phase shift
stabilizes to $31.46$ as $k\rightarrow0$ which is shown in Fig.\ref{li_eta}.

\begin{figure}[h]
   \includegraphics[width=8cm]{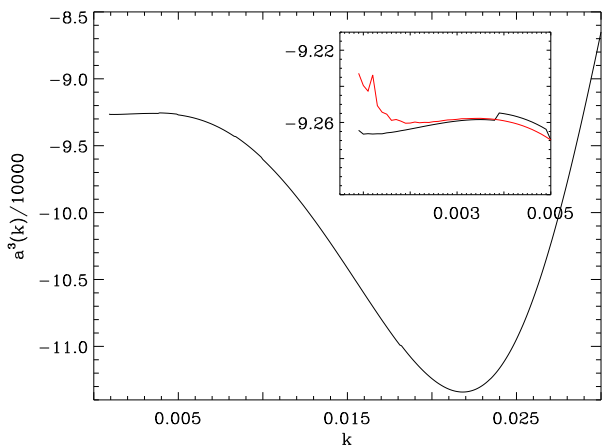}
   \caption{Variation of $-\tan(\eta_1)/k^3$ is plotted as a function of
            $k$. At low $k$ values, the relative tolerance of the calculation
            is $10^{-7}$ . The inset plot shows the sensitivity of 
            $-\tan(\eta_1)/k^3$ to the relative tolerance. The red and black
            curves in the inset correspond to tolerance of $10^{-6}$ and
            $10^{-7}$ respectively.}
\label{li_a}
\end{figure}

 The sensitivity of the phase shift to the relative tolerance at low $k$
is not very prominent from the values of $\eta_1$. However, it is not so with 
the $-\tan(\eta_1)/k^3$, which is the scattering length in the 
$k\rightarrow 0$ limit. The small value of $k$ and the transcendental 
function $\tan$ enhances variations in $\eta_1$. This is evident when 
$-\tan(\eta_1)/k^3$ as a function of $k$ is compared at low $k$ values for 
two different relative tolerances. The plot of such a comparison  is shown 
as the inset plot in Fig.\ref{li_a}. With improved tolerance, the lowest $k$ 
value up to which the reliable phase shift can be calculated is decreased. 
There is however a limitation to decreasing the value of $k$, up to which the 
phase shift is calculated, by decreasing the value of relative tolerance.  
Below a certain value of relative tolerance, the RKF method fails to perform
the integration of the phase equation up to the cut off point. 
From the calculations, we find that $-\tan(\eta_1)/k^3$ stabilizes to 
$\approx -92660$ , this is evident from the plot in Fig.\ref{li_a}). The 
cube root of this is the $a_1$.  Another equivalent way to calculate
the scattering length is to evaluate the slope of $-\tan(\eta_1)/k^2$ 
in the $k\rightarrow 0$ limit. We calculate this by least square fitting 
the data points for $-\tan(\eta_1)/k^2$ between $0.005\leq k\leq0.0005$ with 
an additional point corresponding to zero phase shift at $k=0$. 
\begin{figure}[h]
   \includegraphics[width=8cm]{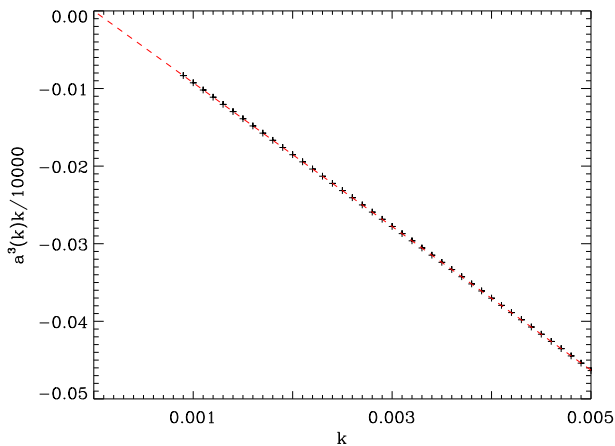}
   \caption{Plot shows least square fitted line (dashed red curve ) on
            $a_1^3(k)k$ values in the domain $0.0005\leq k \leq 0.005$ 
            (black crosses ). An additional point coinciding with origin
             is also considered while drawing the least square fitted curve. 
             The slope of this line is $\approx -92600$.}
\label{lia2}
\end{figure}
As is 
evident from Fig.(\ref{lia2}) there is very good fitting of
straight line over the data points. The slope of the least square fitted 
line is $-92600$, the cube root of which is $-45$ and can be considered as
a reliable value of the $p$-wave scattering length. If we compare it with the 
analytic approximation of
$a_1$ obtained by using Eq~.(\ref{eq.Z.b}) as the value of $A/B$, we find 
that percentage error incurred by using analytic approximation is $43.74$ 
percent.  


\section{conclusions}

For $p$ partial waves, the centrifugal term in the effective interatomic
potential leads to a WKB solution, which is more complicated than the one 
without it as in $s$ partial wave case.  Consequently, the constant $A/B$ 
of the partial wave solution evaluated has radial dependence and apriori,
it is not possible to choose an appropriate $R$ at which a reliable $A/B$
can be evaluated. The approximate analytic expression to calculate $A/B$ 
provide good results for $^{134}$Cs but has large errors for $^{132}$Cs 
and $^6$Li. Based on our calculations, using the variable phase method,
we estimate the $p$-wave scattering length of $^6$Li is $-45$.


\end{document}